\documentclass[sigconf]{acmart}

\usepackage{xargs}
\usepackage{xcolor}
\usepackage{import}
\usepackage{multirow}
\usepackage{booktabs}
\usepackage{todonotes}
\usepackage{graphicx}
\usepackage{pdfpages}
\usepackage{algorithmic}
\usepackage{textcomp}
\usepackage{url}

\usepackage{subcaption}

\acmConference[ICSE 2024]{46th International Conference on Software Engineering}{April 2024}{Lisbon, Portugal}

\AtBeginDocument{%
  \providecommand\BibTeX{{%
    \normalfont B\kern-0.5em{\scshape i\kern-0.25em b}\kern-0.8em\TeX}}}

\setcopyright{acmcopyright}
\copyrightyear{2018}
\acmYear{2018}
\acmDOI{XXXXXXX.XXXXXXX}

\acmConference[Conference acronym 'XX]{Make sure to enter the correct
  conference title from your rights confirmation emai}{June 03--05,
  2018}{Woodstock, NY}
\acmPrice{15.00}
\acmISBN{978-1-4503-XXXX-X/18/06}

\begin{document}

\title{Embedding Sustainability in Software Engineering Curriculum: A Case Study}

\author{Ruzanna Chitchyan}
\email{r.chitchyan@bristol.ac.uk}
\orcid{ 0000-0001-6293-3445}
\affiliation{%
  \institution{School of Computer Science, University of Bristol}
  \streetaddress{Woodland Road}
  \city{Bristol}
  \country{UK}
  \postcode{BS8 1UB}
}

\author{Niki Mahmoudi}
\email{bf25514@bristol.ac.uk}
\affiliation{%
  \institution{School of Computer Science, University of Bristol}
  \streetaddress{Woodland Road}
  \city{Bristol}
  \country{UK}
  \postcode{BS8 1UB}
}
\renewcommand{\shortauthors}{Chitchyan and Mahmoudi}

\begin{abstract}
Sustainability is increasingly recognized as a critical dimension of engineering education, yet its integration into Software Engineering curricula remains a challenge. This paper reports on a case study that examines how sustainability is being embedded across modules in the Software Engineering program at one university. The paper outlines the process through which academics and students co-identified opportunities for integration, guided by the five dimensions of the Sustainability Awareness Framework, targeted discussion questions, and good practice examples drawn from the Green Software Foundation patterns. The study highlights practical steps - including the use of frameworks, illustrative examples, student engagement, and iterative consultative processes - that can support other institutions seeking to embed sustainability into their programs. We also discuss strategies for integrating sustainability into the Software Engineering curriculum and argue that such integration is a necessary and urgent step to prepare Software Engineering graduates as sustainability-aware professionals in our changing society. 
\end{abstract}



\received{20 February 2007}
\received[revised]{12 March 2009}
\received[accepted]{5 June 2009}

\maketitle

\section{Introduction} \label{sec:intro}

Sustainability has emerged as one of the defining challenges of our time, and higher education plays a critical role in preparing graduates to address it. Today it is recognized that engineering disciplines in particular must equip students with both technical knowledge and an awareness of the environmental, social, and ethical implications of their work \cite{moreira2025roadmap,gutierrez_Educational_2025}. The sustainability dimension in engineering encompasses designing and managing sustainable technologies, researching environmental and social impacts and boundaries, and managing resources in a cradle-to-cradle cycle \cite{gutierrez_Educational_2025}. 

Within Software Engineering (SE), these challenges take on a particular urgency. As a discipline that underpins much of today’s digital and societal infrastructure, SE has a responsibility to prepare graduates who can act as sustainability-aware professionals \cite{becker2015sustainability}. However, experience with embedding sustainability into SE education remains limited, and efforts are often confined to superficial curriculum additions or relegated to extra- or co-curricular activities. To help facilitate this process, we share our experience of systematically integrating sustainability into the SE programs at the University of Bristol, UK. 

Using the Sustainability Awareness Framework (SAF) \cite{duboc2020requirements}, which conceptualizes sustainability as five interrelated dimensions (social, environmental, technical, economic, and individual) affected by software solutions, we analyzed how sustainability topics are covered across SE modules. Our methodology combined data collection from:
\begin{itemize}
    \item a survey with responses from 68 students,
    \item analysis of 18 unit specifications and teaching materials,
    \item interviews with 8 academics,
    \item 4 student focus groups, and
    \item a review of open-source sustainability patterns (e.g., those developed by the Green Software Foundation \cite{greenswpatterns2025}).
\end{itemize}

Through the survey, we observed that students wish to better integrate sustainability content into the curriculum (which was previously also noted through the Student Union and its appointed Sustainability Champion in the School of Computer Science). Thereafter, through interviews and focus groups, the teaching staff and student participants explored specific units, considering the key research question: \textit{How and where should sustainability be integrated into our SE curriculum?}
To address this question, we explored if and where sustainability is already present in our units, what relevant concepts are currently missing, and how they might be taught and assessed. 

Below we report that the structured discussion sessions — supported by SAF dimensions, guiding questions, and sustainability patterns — helped academic colleagues identify relevant opportunities for integrating sustainability into their modules. These sessions also served as a platform for co-developing concrete next steps toward such integration by curating example patterns for good software engineering practice from the Green Software Foundation \cite{greenswpatterns2025} and agreeing on a mechanism for iterative progress review.

The contribution of this paper is therefore twofold: (1) to present the current state of sustainability integration across the SE curriculum at \textit{ University of Bristol}, and (2) to share practical steps that may help other institutions replicate and advance such integration efforts. With this, we hope to contribute evidence of good practice in embedding sustainability into SE education and provide a simple method that can be replicated across other higher education institutions.

Finally, we discuss our experience against the sustainability integration strategies typology suggested by Kolmos et al. \cite{kolmos2016response}. Kolmos and colleagues distinguish between add-on strategies (small, visible changes such as new modules added to existing programs), integration strategies (embedding sustainability across curricula), and rebuilding strategies (redefining disciplinary boundaries). We observe that both add-on and integration strategies are present in the University of Bristol case study. While the rebuilding strategy has yet to emerge, we suggest that the integration strategy is both a necessary prerequisite to it and an urgently needed step for the SE discipline today. 

The remainder of this paper provides an overview of related work (see Section~\ref{sec:relatedWork}), presents our study methodology more formally (see Section~\ref{sec:methodology}), outlines the key findings from our data analysis (see Section~\ref{sec:findings}), and discusses the broader implications and lessons learned in Section~\ref{sec:discusison}. The study concludes with a call to undertake a concerted effort to roll out the integration strategy across a wider set of higher education SE institutions (Section~\ref{sec:conclusion}).

\section{Related Work}\label{sec:relatedWork}

Related work on how sustainability is integrated into Software Engineering teaching and learning tends to fall into three categories: (a) general curriculum design, (b) skills and competencies, and (c) demonstrations of how specific sustainability topics (such as energy efficiency) are addressed. Each of these categories is briefly discussed below.   

\subsection{Curriculum Design}\label{sec:litMethods}
Work on curriculum design is often situated in the broader engineering domain, of which Software Engineering is one instance. Here, research considers the composition of sustainability content and ways of integrating it into (software) engineering teaching.
For instance, Fisher et al.\ \cite{fisher_Incorporating_2016} suggest that such integration can be undertaken at the \textit{course level}, \textit{course-component level}, or \textit{curriculum level}.
At the \textit{course level}, integration occurs by introducing modules explicitly focused on sustainability-related topics. Examples include \textit{Computing, Energy, and the Environment} and \textit{Algorithms for Ecology and Conservation}. These modules demonstrate how computational methods address real-world sustainability challenges. At the \textit{curriculum level}, a specialized track may be developed, combining specific prerequisite courses (e.g., Data Structures) with later sustainability topics (e.g., \textit{Efficient Solutions for Hydrology}). Within such a track, modules would be differentiated by the sustainability domains in which students specialize.
Finally, at the \textit{course-component level}, sustainability themes can be embedded directly into lectures, exercises, and projects within existing Software Engineering modules. This approach enables students to engage with sustainability issues even when enrolled in general Software Engineering tracks, ensuring broad exposure beyond specialized modules. 
While Fisher et al.\ \cite{fisher_Incorporating_2016} account for multiple levels of integration, they fail to address core competences needed for developing sustainably engineering skills in CS students. Instead, they focus on how computational tools and methods can support other disciplines (such as hydrology or ecology) in addressing sustainability concerns. 

Boyle et al.\ \cite{boyle2004considerations} consider the \textit{student's readiness} to grasp the challenges of integrating sustainability into engineering. They note that, given the complex, systemic, and dynamic nature of sustainability, \textit{``Most students entering post-secondary education \dots have a limited
understanding of or interest in them [i.e., systemic and complex topics]. Such students are highly focused on their own personal needs and have difficulty determining how they fit into society and
how society functions. [\dots] Consequently, first-year students have difficulty comprehending
the applicability of sustainability engineering \dots''}. Thus, Boyle et al.\ \cite{boyle2004considerations} suggest teaching sustainability either as a master’s-level program or as one additional unit taught per year to a selected set of ``bright'' students. 

A set of related studies demonstrates that universities across the world (e.g., from Australia \cite{arefin2021incorporating}, to Sweden \cite{leifler2020curriculum}, and Brazil \cite{sigahi2023isolated}) are all interested in integrating sustainability content into their educational provision. Yet, all this work still remains within the integration levels proposed by Fisher et al.\ \cite{fisher_Incorporating_2016}. 

\subsection{Generic Skills, Competencies, and Challenges} \label{sec:litSkills}
The desire to cultivate sustainability-related skills in Software Engineering and Computer Science professionals stems from both industry \cite{heldal2024sustainability} and academia \cite{leifler2020curriculum,moreira2024road,chitchyan2016sustainability}. Related work that examined software engineers' sustainability-related skills and their relevance in practice is primarily based on practitioner surveys and interviews \cite{karita2021software,betz2022software,chitchyan2016sustainability,heldal2024sustainability}.
Earlier papers indicated that practitioners considered sustainability to be of little professional relevance to themselves \cite{groher2017sustainability,chitchyan2016sustainability}. Where they did see sustainability as relevant, it was mainly with respect to technical and quality attributes. This stance has not changed much in recent years; although ICT organizations now consider sustainability to be substantially relevant to them \cite{karita2021software}, they still distinguish 
\textit{``IT people from sustainability experts''} \cite{heldal2024sustainability}. The sustainability skills asked of
software engineers are still centered on technical topics such as software quality, user-centric design, data management, and architecture. Yet, skills such as systems thinking, programming for energy efficiency, communication skills, and advanced data management \cite{heldal2024sustainability} are now considered necessary but are often expected from ``sustainability professionals.'' 

\subsection{Addressing Specific Sustainability Topics}\label{sec:litSpecific}
A number of studies have considered how to teach various sustainability-specific topics within individual Software Engineering modules. For instance, Duboc et al.\ \cite{duboc2020requirements}
discuss how to integrate awareness of sustainability into a Requirements Engineering module using a set of directed questions related to various aspects of environmental, societal, technical, economic, and individual sustainability. They also advocate for consideration of time-related effects, such as the long-term use of a given software by many users at a time. 

Others focus on integrating sustainability into the agile development process \cite{bambazek2023requirements,oyedeji2024integrating}, addressing sustainability-related risk in 
architecture design \cite{ojameruaye2016sustainability}, or minimizing the energy consumption of code generated by large language models through prompt modification \cite{cappendijk2024generating}.

Equally notable are the efforts of practitioners to collect and disseminate skills and experiences related to efficiency in energy, water, and land-use activities in Software Engineering \cite{greenswpatterns2025}. 

All three categories of related work are essential in developing a Software Engineering curriculum that integrates sustainability. Section~\ref{sec:litMethods} outlines how curricula could be structured, work on skills and challenges (Section~\ref{sec:litSkills}) highlights which skills are essential and which challenges need to be addressed, and work on specific solutions (Section~\ref{sec:litSpecific}) demonstrates experiences of integrating individual aspects. Yet, our community still lacks a clear example demonstrating how such a curriculum can be derived. This challenge is addressed in the rest of this paper. 

\section{Methodology} \label{sec:methodology}



\subsection{Research Design}
This study adopted an \textit{action research} methodology to investigate and support the integration of a sustainability into the computing curriculum of the School of Computer Science at the University of Bristol. Action Research was selected for its emphasis on iterative, participatory change and its suitability for educational contexts where practitioners and researchers collaborate to improve practice \cite{efron2019action,kemmis1988action}. 

The research was conducted within the School of Computer Science at a UK university, with the \textit{participants} comprising academic staff members responsible for teaching undergraduate and postgraduate modules, as well as students currently studying at the School. All academic staff and students were invited to participate. All participation was voluntary. Each academic staff participant engaged individually, while students were invited to participate in group discussions. All participants provided informed consent in accordance with institutional ethical guidelines. The list of participants is provided in Table \ref{tab:participants}. 
\begin{table}[h]
\centering
\caption{Participants: Interviewees (P ID) and Focus Groups (FG ID).
}
\begin{tabular}{|p{4cm}|p{0.5cm}|p{0.8cm}|p{1cm}|}
\hline
\textbf{Unit Name (Code)} & \textbf{Year} & \textbf{P ID} & \textbf{FG ID} \\ \hline
Applied Deep Learning (ADL) & 3 & P4, P8 &  F4\\ \hline
Computer Architecture (CA) & 1,4 &  & F3, F2 \\ \hline
Computer Systems (CSys) & 2 & P4 & F4, F1 \\ \hline
Cyber Security (CSec) & 3, 4 & P3 &  \\ \hline
Applied Data Science (DS) &  3& P5 &  \\ \hline
Foundations of Cyber Security (FCS) & 4 & P3 &  \\ \hline
Imperative and Functional Programming (IP) & 1 & P8 & F3 \\ \hline
Interaction and Society (InS) & 2 & P2, P6 & F4 \\ \hline
Interactive Devices (ID) & 3,4 & P1, P7 & F4 \\ \hline
Human-Computer Interaction (HCI) & 3,4 & P2, P6 &  \\ \hline
Machine Learning (ML) & 3 & P4, P5 & F1 \\ \hline
Mathematics and Statistics (MnS) & 1 & P1 & F3 \\ \hline
Object Oriented Programming and Algorithms (OOP) & 1,4 & P8 &  \\ \hline
Overview of Software Tools (OST) & 1,4 &  & F3, F2 \\ \hline
Programming in C (PiC) & 4 & P8 & F2 \\ \hline
Security Behaviours (SB) & 3 &  P3 & F1 \\ \hline
Software Engineering (SE) & 2,4 & P7 &  \\ \hline
Sustainable Computing (SC) & 4 &  &  \\ \hline
\end{tabular}
\label{tab:participants}
\end{table}

The \textit{intervention} consisted of an interview or focus group, using a previously well publicized sustainability framework (SAF) \cite{duboc2020requirements} outlining key principles and concepts related to sustainability (e.g., ethical risk, responsible innovation, human-centered design, environmental impact of software and hardware, etc.). The framework aimed to support educators in identifying relevant concepts for their units and embedding these concepts into their teaching and assessment practices.
The research \textit{procedure} comprised two cycles:
\begin{itemize}
    \item \textit{Introduction and diagnosis cycle}: we first carried out a student survey to elicit the perceptions of the student body on the current state and need for further integration of sustainability into the taught program. Then each participating academic staff member was introduced individually to the sustainability framework in a one-on-one meeting, while the same introduction was given to the student groups of 3 to 6, before the discussions. During these meetings, participants were asked to describe if and how they currently taught/were taught the related sustainability concepts within their modules. These conversations served both to raise awareness and to surface existing practices and perceptions with both staff and students.
    \item \textit{Collaborative planning cycle}: Participants were then invited to consider how they might integrate elements of the sustainability framework into their teaching /taught content in upcoming iterations of the respective modules. Where relevant, the researcher suggested concrete examples tailored to the module context (e.g., case studies, student activities, assessment modifications). These discussions were documented and  tentative integration steps outlined (with staff and students) and agreed upon (with staff). After the meeting the agreed plan of actions was emailed to the participant staff members.
\end{itemize}
It was worth noting that as part of interviews it was agreed that (in approximately one year) a follow-up reflection discussions will be held with each interview participant to review progress on addressing the set plans. These follow-up interviews are part of our future work and are not discussed further in this paper. 

\subsection{Data Collection and Analysis}
To support the present study we used two artifacts: Sustainability Awareness Framework \cite{duboc2020requirements}, and 
Sustainability Patterns from Green Software Foundation \cite{greenswpatterns2025}.

\subsubsection{Supporting Artifacts}
To support the interview and focus group discussions, we utilized the Sustainability Awareness Framework \cite{duboc2020requirements} which  introduced participants to sustainability through five key dimensions: Social, Individual, Environmental, Economic, and Technical. In this framework each dimension is accompanied by a set of topics and questions to prompt reflection and guide conversation. For example, the Social dimension covered issues such as sense of community, inclusiveness, diversity, and communication, while the Environmental dimension highlights use of materials and resources, pollution, energy, and biodiversity. The table of dimensions and topics (see Table \ref{tab:sustainability_dimensions}) was presented alongside a visual diagram to help participants quickly grasp the breadth of sustainability considerations and consider their relevance to software and its engineering.
\begin{table}[h]
\centering
\caption{Topics Covered per Modules and Sustainability Dimensions}
\label{tab:sustainability_dimensions}
\begin{tabular}{p{0.088\textwidth} p{0.35\textwidth}}
\toprule
\textbf{Dimension} & \textbf{Topics} \\
\midrule
Social & Sense of community; Trust; Inclusiveness \& Diversity; Equality; Participation \& Communication \\
\midrule
Individual & Health; Lifelong learning; Privacy; Safety; Agency \\
\midrule
Environmental & Materials and Resources; Soil, Atmospheric and Water Pollution; Energy; Biodiversity and Land Use; Logistics and Transportation \\
\midrule
Economic & Value; Customer Relationship Management; Supply chain; Governance and Processes; Innovation and R\&D \\
\midrule
Technical & Maintainability; Usability; Extensibility and Adaptability; Security; Scalability \\
\bottomrule
\end{tabular}
\end{table}

Green Software Foundation is an open source initiative aimed at collecting and sharing the good practices on environmentally-aware software development and deployment - the Sustainability Patterns \cite{greenswpatterns2025}. These patterns vary across development contexts (e.g., use of AI, web development, etc.), but provide a grounded appreciation on the environmental costs of engineering decisions (e.g., impact of linking up sizeable images into the web design).
In preparation for each interview/focus group, the researchers reviewed the patterns and identified those relevant to each module. These were brought to the interviews/focus groups to illustrate possible technical solutions and skills that can be taught within the relevant modules. 

\subsubsection{Study Procedure}
The student survey was carried out during the Computer Science Society industry event. All CS students are members of this society and can join all it's events. A brief survey asking if students felt that the current program already covered and should include more topics on sustainability was designed and administered by the Student Sustainability Champion during one of the meetings. The survey started with presenting the Sustainability Awareness Framework to explain what it meant by `sustainability', so as the students responded with the same conceptual reference in mind.  

Interviews and focus groups were overpraised post-survey. All interviews and focus groups started with an introduction to the study, explaining its purpose of embedding sustainability into the curriculum of the School, and presenting the Sustainability Awareness Framework to explain what we mean by `sustainability'. 
Thereafter, the study participants were invited to consider each module they have been studying/teaching and discuss:
\begin{itemize}
    \item Introduction and Diagnosis Questions: 
\begin{itemize}
    \item What content covered in a given module related to sustainability?
    \item What sustainability-related concepts (if any) should be covered in a given module (but have not been so far)?
    \item How should these concepts be integrated/taught (e.g., via case studies, as part of the labs, in any other way)?
    \item How should the integrated concepts be assessed (if at all)? 
\end{itemize}
\item Collaborative Planning Questions (only used for interviews)
\begin{itemize}
    \item How can we best integrate the previously discussed sustainability content into this module?
    \item Do you think the given sample of Green Software Foundation's Sustainability Patterns is relevant to your module? If so, how can the respective patterns be integrated into the teaching and learning activities of the module?
    \item What actions should we agree to for integrating sustainability into the given module? 
\end{itemize}
\end{itemize}
The list of final agreed upon actions was emailed to each interview study participant 1 to 3 weeks post interview. 

\subsubsection{Data Collection and Analysis}
Data from \textbf{student survey} (completed by 68 respondents) was collected via a Google form and analyzed for summary statistics.

Data collection for \textbf{focus groups} (4 groups of 3 to 6 members each, with 19 participants overall) was carried out through field notes taken by two researchers, wile \textbf{interviews} (with 8 participants) were recorded and transcribed.  
In preparation for the meetings, the respective module pages and teaching plans where shared with the researchers and reviewed as well. 

All data were anonymized and stored securely. 
Data from field notes and focus groups
were analyzed using thematic analysis \cite{braun2006using} to identify currently covered and missing sustainability-related content, modes of such content  delivery and assessment, barriers and enablers to integrating it into teaching and learning. A deductive coding approach was used initially, based on anticipated themes such as currently addressed content, missing content, and assessment methods, followed by inductive refinement based on emerging insights.

The coding of one transcript was simultaneously and independently carried out by two researchers with results discussed and harmonized for this sample transcript to ensure consistent category identification. Thereafter the analysis was carried out by one of the researchers and reviewed by the other. We did not aim to arrive to any saturation results at this stage of the study, as each module was expected cover rather different content, and we aimed to review as many modules as we could cover within the available time, and continue the process in the following year.  

\section{Findings} \label{sec:findings}
\subsection{Student Survey Results}\label{sec:studentSurveyResults}
\begin{figure*}[htb]
    \centering
    \includegraphics[width=0.8\linewidth]{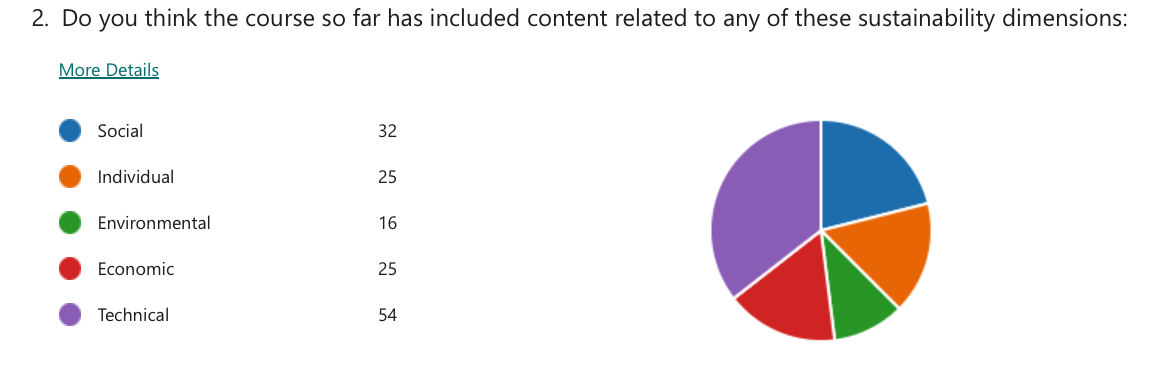}
    \caption{Student Perceptions on Sustainability Covered in their Program}
    \label{fig:studentCurrentCovered}
\end{figure*}
\begin{figure*}[htb]
    \centering
    \includegraphics[width=0.8\textwidth]{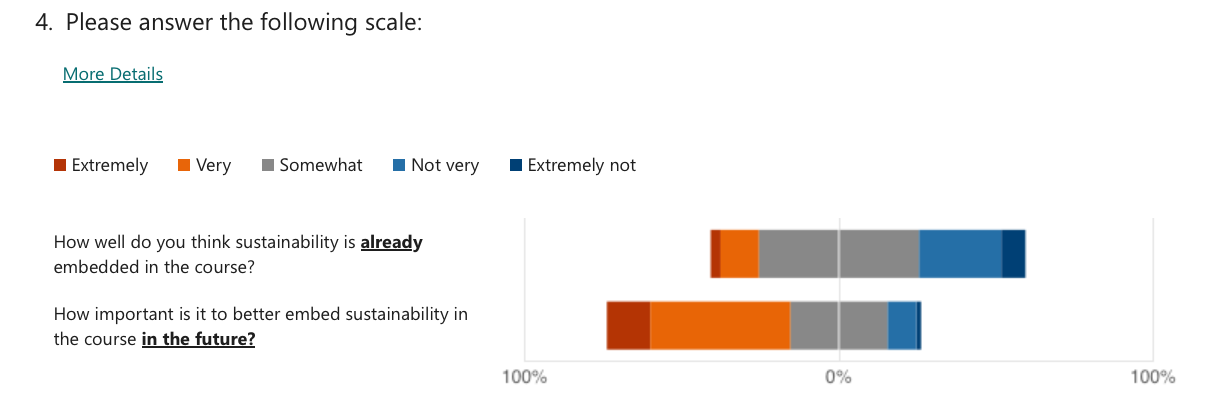}
    \caption{Student Perceptions on Current and Desired Embedding of Sustainability in CS Program}
    \label{fig:sustEmbedding}
\end{figure*}
The survey results on students' perceptions of the extent to which their course so far has included content related to different sustainability dimensions are shown in Figure~\ref{fig:studentCurrentCovered}. The responses suggest that the technical dimension is most strongly represented, with 54 mentions. This is followed by the social dimension (32 mentions), while the individual and economic dimensions were identified less frequently, each with 25 mentions. The environmental dimension received the lowest recognition, with only 16 mentions. The accompanying pie chart highlights the dominance of technical content and the relative underrepresentation of environmental and individual/economic aspects.

Figure~\ref{fig:sustEmbedding} shows students' perceptions of the extent to which sustainability is already embedded in their course, compared with how important they feel it is to embed sustainability in the future.

The first bar indicates that current sustainability coverage is generally viewed as limited. Most students rated it as only somewhat embedded (50.8\%), while a substantial proportion considered it not very (26.2\%) or extremely not embedded (7.7\%). Very few selected the extremely (3.1\%) or very well (12.3\%) embedded categories.

By contrast, the second bar highlights a strong consensus on the importance of future integration. Here, most responses fall into the extremely (13.8\%), very (44.6\%), and somewhat important (30.8\%) categories.

Taken together, the results point to a clear gap: while students perceive current sustainability content as weak, they place high importance on embedding sustainability more strongly in the course going forward.

\subsection{Academics' Perceptions and Reflections}
Academics' opinions varied about the sustainability dimensions before and after the interviews. 

\subsubsection{Pre-discussion Viewpoint (Perspectives)}
In many cases, the academics admitted that they lack the required sustainability awareness to integrate related material in their units, e.g., \textit{"I want to say first a like I have a very narrow definition of sustainability." [P1]; "I have been teaching this unit and it has never crossed my mind that I am looking at all of this sort of sustainability principles, just teaching it." [P3]; "[...] I am not knowledgeable about this side of it." [P4]; "The lecturers need to be trained in the topics to be able to integrate [sustainability]." [F4]}. Academics lack knowledge about relevant sustainability resources, which is one of the contributing factors to the lack of sustainability awareness. In some cases, academics have not prioritized sustainability as they have mainly considered the theoretical aspect of teaching, e.g., \textit{"I have only kind of thought about my part of the unit for a purely theoretical perspective." [P4]}. 
Some of the stigma around sustainability is that when it is mentioned people instantly think of reducing waste; this also indicates a lack of sustainability awareness. 

\subsubsection{Post-discussion Viewpoint (Reflections)}
There was a shift in academics' perspectives after the interview. Post-discussion, academics acknowledged the sustainability problems, e.g., \textit{"I did not know about the Green Software Foundation \dots" [P8]}. The academic staff showed interest and were keen on making changes toward developing a more sustainable framework for students. It was proposed that sustainability should be introduced as a non-optional module, given its importance: \textit{"[Sustainability] is competing with many other choices. So this needs to be integrated into units that are not optional" [F4]}. It was generally agreed that a good representation for sustainability material is needed.

As part of the discussion process, we observe that starting with the SAF framework was a particularly relevant educational tool that broke down a number of misperceptions and related resistance to the relevance of sustainability to a given unit. Similarly, using the GSF patterns was invaluable in demonstrating specific solutions handy for teaching in each unit. 

\subsection{Unit Content Analysis}\label{sec:subjectAreaAnalysis}
The existing teaching material addresses sustainability implicitly in different ways and dimensions specific to the unit itself. We discuss the units and their treatment of sustainability in related categories based on specific SE aspects they address. These are: 
\begin{enumerate}
    \item Programming category units, including Programming in C, Imperative and Functional Programming, and Object-Oriented Programming and Algorithms units;
    \item Cybersecurity category units, including Foundations of Cybersecurity, Security Behaviors, Cybersecurity;
    \item Human-Computer Interaction category units, including Interaction and Society, Interactive Devices, Human-Computer Interaction;
    \item Systematic Software Engineering category units, including Software Engineering, Overview of Software Tools, and Computer Systems;
    \item Hardware category units, including Computer Architecture, and Interactive Devices (hardware aspect of it);
    \item Intelligent Systems category units, including Mathematics and Statistics, Machine Learning, Artificial Intelligence, and Advanced Deep Learning units;
    \item Responsible Computing category, which currently includes only one unit—Sustainable Computing.
\end{enumerate}

\subsubsection{Programming Units}
The programming units focus on logic and understanding the capabilities, syntax, and semantics of the programming language. Through Programming in C (PiC), students understand how memory management works, e.g., \textit{"The C language often relates to the workings of a machine, It is so close to the Von Neumann Architecture in terms of understanding how computation operates." [P8, IP]}. Space and time complexity are a part of the currently covered material that help students understand the most suitable application for each programming language, e.g., \textit{"We say C++ or C is very good compared to Python, or C is better than Java because it takes less time for instance or in algorithms." [P5, IP]}. These concepts demonstrate the basics of programming languages and implicitly link these to resource management (memory, time, complexity), all of which will have impacts on the environment and human society, but this aspect remains omitted from the module. 

Students later undertake the Object-Oriented Programming and Algorithms (OOP) unit where they practice programming with objects and learn about implementing maintainability, usability, adaptability, and security into their code. OOP and its key principles—from the very basics (e.g., no repetition in code), to promoting use of good-practice design patterns, use and reuse of software components, and moving from language-specific software to language-agnostic software components—promote optimization of resource use (often focusing on compute, time, and money). OOP also introduces the processes and principles of development practices (e.g., Agile development), which relate to the topics of technical excellence and economic efficiency. 

Students first learn of the need for correctness in developing software through working on a test-driven framework. All unit test frameworks are open-source and transparent to strengthen the social (with respect to the development team) aspect of sustainability. The need for code to be maintainable and readable by other programmers is also discussed to highlight possible social impacts (for the clients and broader community), e.g., \textit{"This is the basis for the software crisis. Software is late, not on budget, not on time... Satellites fall out of the sky because of it" [P8, IP].}

Resource allocation and its starting points such as data structures (for dynamic allocation), memory allocation, and computational models are all covered within these units, but they can also be linked to the notions of sustainability by connecting these to the limits of processing power and the economic impacts of programming.

Although \textit{"[The programming unit] currently does not consider how much memory the solution is using. At the moment we have abundance of memory and no single programmer needs to think about it." [F3, IP]}, through the focus group discussion, the students recommend that the units integrate conceptual notions of environmental impacts of the current programming tasks.
Echoing the same ideas, interviewee P8 notes for the context of the PiC unit: 
\textit{``You [programmer] are in charge[...] and there is potential to integrate a little bit of that [i.e., sustainability] from the very beginning of the pipeline by explicitly linking it to resource utilization."}

As a result of the discussions, the IP unit has planned to explain good coding practices and then create a clearer link from these to the impacts of programming, e.g., \textit{"It is more important to get the basics properly sorted and we can improve the unit by clearly outlining why you are doing it [implementing good programming practices] and what the repercussions are if you do not" [P8, IP].} The PiC unit intends to include introductory lecture slides to introduce students to the different sustainability dimensions. As noted by P8: 
\textit{``You are going to learn how to turn your brain power into providing programmes which have a social, individual, environmental, economic, and technical dimensions. This is the beginning of your journey, and wherever you are on your journey, remember this" [P8, PiC].}

\subsubsection{Cybersecurity Units}
Cybersecurity (CSec) currently covers such social aspects of sustainability as inclusiveness in supply chains and providing security solutions that create a sense of community and trust while protecting all social groups.  

In cybersecurity, the topic of trust is crucial as it underlies the design of threat models, e.g., \textit{" \dots it is an important principle in terms of making sure that people trust the mechanisms that are there, because if they don't trust them, they don't use them" [P3, FCS]}. Thus, Foundations of Cybersecurity (FCS) teaches the relationships between systems and collaborative trust among them, as well as such topics as usability, adaptability, security, and cryptography. For instance, it explains why usable and accessible authentication mechanisms are needed, how cryptography is used in blockchains to have immutable components, and how security breaches could cause harms, e.g., overrunning data centers with malicious requests which could result in environmental problems. This unit also addresses some aspects of individual sustainability (such as agency, safety, and privacy) by designing systems that users feel confident to take away and use for themselves. 

The FCS unit also discusses economic sustainability in supply chain, governance, and processes by looking at security operation centers and touches on the topics of vulnerabilities and their economic influences.
It also includes technical sustainability topics such as co-designing usable, scalable, and maintainable systems that allow other developers to build on them and perform security updates, as well as developing scalable security mechanisms.

Security Behaviors (SB) includes materials that help students understand a user's perspective in addition to the software system's. Biases (such as hand soap dispensers not recognizing dark skin, biases in face recognition, etc.) and how to design to avoid them are also addressed. As noted by P3: \textit{``When we are discussing issues around security or privacy, we should not just think of ourselves as individuals. We should also think about other groups that may be affected by the mechanisms or the controls that we design."} For instance, the unit looks at the challenge of designing an inclusive multi-factor authentication service. While such authentication services (generally designed around smartphone use) are widely rolled out today, little to no consideration is given to minority groups that might not be able to access smartphones. They are left with no alternative access to digital services.

The study participants noted that security breaches can have serious negative environmental impacts if, for instance, botnet attacks disrupt digital services that run manufacturing and other chemical/biological processes. Thus, though this is not addressed in the current set of cybersecurity units, in the future it would be relevant to consider how such attacks on essential services (e.g., healthcare) affect both people and the environment.

Following the discussions, it was agreed that the environmental aspect of the units would be improved through a more explicit reference to sustainability via a dedicated case study. As proposed by students in focus group F4: \textit{"[One way to integrate this could be] if you present a badly designed example and ask them [students] to reflect what is wrong with it?" [F4]}. Students also wanted lecture slides and questions that address the sustainability dimensions as part of the teaching materials. Participant P3 also suggested that round-table discussions about the topic of sustainability and cybersecurity can encourage students to think about their impacts, and students will have the opportunity to present the discussion results for their tables to the whole class.

\subsubsection{Human-Computer Interaction}
The Interaction and Society (InS) unit currently covers teaching material related to values, accessibility, decolonization, ethics, law and order, and war and peace (e.g., technology used in policing and warfare). The students are also informed of the Computing Profession Requirements from the 
BCS Chartered Institute for IT (the CS  Accreditation Body in the UK) — all of which relates to social sustainability concerns. Students then continue to the Human-Computer Interaction (HCI) unit where they learn to comprehend and appreciate multidisciplinary human-computer interaction by recognizing users' perspectives, e.g., \textit{"You are not your user.
Different perspectives on technology may exist. [The unit explains] some simple methods that other people might be using to help you design technology to fit with human beings a bit better." [P6, HCI]}

The Interactive Devices (ID) unit then makes students engage with actual end users, deepening the link to the individual and social dimensions of sustainability. Innovation, research, and development, which is part of the economic sustainability dimension, is covered through the innovative characteristic of the ID unit by inviting students to design a device that is their own original creation. Students learn the basics of responsible innovation, as noted by P1: 
\textit{"So [\dots] they [students] have to dig a little bit about some aspect of responsible innovation and the impact of it."}

As a result of our discussions, the colleagues noted that the current InS unit is lacking content on systemic perspectives and is only focused on the value conflicts among economic, social, and individual pressures. In addition, the HCI unit would need to embed content related to the environmental aspects of sustainability (e.g., given that some color schemes are more energy-intensive than others, even the choice of color in web design can be environmentally significant).
As for the ID unit, it could set a challenge to design a sustainability-related device or project that can highlight the importance of this dimension as well as promote creative sustainability-focused problem solving. 

\subsubsection{Systematic Software Engineering}
The Software Engineering (SE) unit focuses on such technical sustainability aspects as maintainability, usability, and design reuse. Designing distributed systems and their benefits, drawbacks, and costs, such as time and space requirements, are reviewed throughout both units. More specific topics of distribution, concurrency, and parallel processing are currently covered in Computer Systems (CSys) but have no reference to their respective impacts on sustainability (e.g., in terms of metals and material consumption for hardware manufacturing or energy consumption for processing). While the units touch upon these distributed systems’ local caching costs and compare them to a cloud-hosted environment, recognizing the importance of transfer costs, the focus group participants underlined that while the speed and costs of cloud-hosted systems are compared, their environmental impacts, or the impact of data transfer, are not addressed. 

Topics related to Big O notation and consensus in distributed systems in the presence of faulty nodes are already discussed with respect to energy consumption, which relates to environmental sustainability for the CSys unit. On the technical side, students are introduced to algorithms with attention to runtime and space limitations (though with no explicit mention of sustainability), as well as principles of usability, maintainability, and opportunities for reuse. Students also gain experience in benchmarking their programs to analyze performance and evaluate code quality—all of which contributes to the technical sustainability agenda.

CSys addresses individual sustainability through discussions of privacy and protection levels, particularly in relation to how these issues evolve when systems are scaled up to larger infrastructures and multiple machines. Data collection is also linked to the social dimension of sustainability within this unit, though only implicitly. As one participant noted in a focus group:  \textit{"Scalability is being taught as a part of the functionalities and not under sustainable practices" [F1, CSys].} 

The Overview of Software Tools (OST) unit introduces different communication tools and development platforms (e.g., Git, Teams) which help to create a sense of community for software engineering teams. Git introduces students to a secure working environment where maintainability and usability are fostered, highlighting the technical sustainability aspects of SE. Git security measures and use practices (e.g., on sharing GitHub repositories) are also taught. 
Focus group (F2) participants also suggested including topics on database optimization (e.g., normalization for storage resource reduction) as part of the materials with impact on environmental sustainability. 

As a result of our discussions, the academic delivering the SE unit intends to include open-ended exam questions in the unit assessment where marks will be awarded for incorporating sustainability considerations into answers. The unit will also seek to connect sustainability with established good practices, as the necessary foundations are already in place. They will also consider introducing a ``sustainability stamp'' for software-related topics in all unit exercises, both to recognize and to encourage students' engagement with the sustainability impacts of their work. The CSys unit director plans to review their lectures to frame such themes as privacy, inclusiveness, and trust as explicitly related to sustainability, as they noted: \textit{``These [are] material that we want to talk about especially, but we have not grouped them as sustainability''} [P4, CSys].

\subsubsection{Hardware-Related Topics}
Hardware-related topics such as Computer Architecture (CA) and Interactive Devices (ID) are analyzed in relation to how the software runs. ID focuses on device usability and maintainability—all topics of the technical dimension of sustainability. It also introduces accessibility and end-user engagement to motivate students to develop more socially sustainable devices.
Memory management and garbage collection are taught here too, but do not explicitly mention energy consumption of the hardware, as the key focus is on how the technology works. 

Students are encouraged to use reusable battery packs in the hardware components of these units to minimize negative environmental impacts. For accessibility, they are also required to design devices that are simple to produce and use inexpensive materials. In addition, students are encouraged to reuse materials for prototypes when working with 3D printers, although this practice has yet to be formalized into explicit guidance. Proper and efficient operation of laboratory machinery is another area that requires greater emphasis.

As part of ID, students are also encouraged to collaborate and communicate in groups to empower their social sense of trust, diversity, and inclusiveness (for the development team members).

As a result of our discussions with students from focus groups (F2 and F3) and colleagues (P1, P7), it was agreed that in ID and CA the use of cheap and accessible materials (e.g., for creating transistors in CA or various devices in ID) is already covered, but without being explicitly framed in terms of sustainability; this will be reframed. Likewise, discussions of cost and energy consumption in device production could be extended to highlight sustainability concerns, connecting these topics to both the economic and environmental dimensions of sustainability.

\subsubsection{Intelligent Systems Design}
Mathematics and Statistics (MnS) are the basic starting points for this cohort of units. MnS teaches students to test and evaluate experimental designs, with R used for visualizations. The ethical implications for statistical analysis—such as participant consent and approval, data privacy, and considering the impacts of the technology—are taught, linking to but not explicitly naming the social and individual dimensions of sustainability. 

The Advanced Deep Learning (ADL) unit explores the social dimension of sustainability through ethics and privacy. Non-representative datasets and resulting model biases—with their impact on diversity and inclusiveness—are discussed. However, these topics are \textit{``not [counted] towards any real learning objectives or outcomes" [P4, ADL]}. Training and optimization techniques (e.g., data augmentation toward diversity) are a relevant part of the social sustainability dimension and are already covered in this unit.

Practical labs cover essential topics such as basic classification exercises and topics such as how ChatGPT operates, its inference pass, and how economically costly it is in terms of energy consumption. The impact of tokenization problems and the cycling of ChatGPT is also discussed to explain how the outputs are generated and financial costs accrued.

The ADL unit ends its journey with students reaching the latest Transformer systems. 
And while the teaching staff (P4, P8) are well aware that AI systems require large amounts of water and energy resources—with these trends and the respective environmental impacts set to rise due to the rapid growth of these technologies \cite{bolon2024review}—they have not yet integrated the environmental impact issues into the unit's formal teaching and assessment content.

The Applied Data Science (DS) unit already addresses several sustainability dimensions, though none of these are explicitly mentioned. For instance, the social sustainability dimensions are addressed through themes of diversity, inclusiveness, participation, and communication, particularly when students engage with diverse datasets. The individual dimension is touched upon for students themselves as they work collaboratively in small teams where fair contribution is expected. Privacy is introduced through topics such as algorithmic bias, transparency in white-box and black-box models, and the safeguarding of sensitive attributes to ensure data security and safety.

However, the environmental and economic dimensions of sustainability are largely absent from DS. Issues such as the energy consumption of data processing, the material impacts of large-scale computation, or the economic implications of algorithm deployment are not explicitly addressed.

In the student-led focus groups, algorithm efficiency and model selection were discussed. Students also noted that the units contain no mentions of energy use and impact on the planet when these topics are taught. As per the students, currently, training model biases and choosing the right training models are not well addressed within the CS program.

Energy needed to train Machine Learning (ML) models and techniques to make models as efficient as possible are all material that students noted as relevant to add to these modules. Adding new examples (and possibly research) ideas for students—such as ML in health and biodiversity or neuroscience—can be helpful for them to understand the effects these models have.

As a result of our discussions, the DS unit is planning to broaden the definition of ethical implications by bringing in sustainability themes, as noted by P5: \textit{``We give some way to ethics right now, if you make it ethics plus sustainability implications, it goes directly in the mark scheme in that case."} The sustainability dimensions are also planned to be added to the DS data privacy materials. Sustainability will be integrated in the form of a requirement in the ADL unit, e.g., \textit{"The more and more it becomes actually a technical requirement to even be able to train your system to be resource efficient. It is not just that we are nice to the environment [...] It is literally impossible to train it otherwise because it will take three months" [P8, ADL]}. Some of the ways to deliver sustainable solutions will be incorporated into teaching tasks themselves, e.g., designing algorithms that are less compute intensive and finding ways of using smaller models.

\subsubsection{Responsible Computing}
The Sustainable Computing (SC) unit of the SE program is explicitly dedicated to teaching responsible, sustainability-focused software engineering. On the technical side of SC, students learn to identify and consider the potential harms and impacts of their software solutions, and consider possible alternatives (e.g., making online music streaming platforms more energy efficient by changing the format of music files that are stored and streamed).
Value conflicts and societal pressures are identified and discussed to demonstrate the tensions in environmental vs. economic vs. social and individual impacts of software development. The unit also teaches core approaches to technology life-cycle accounting, considering the environmental impacts of cradle-to-grave technology for materials and resource use. This, however, is a single unit, and does not have the scope to cover social, economic, and individual impacts of technology in sufficient detail. 

\section{Discussion}\label{sec:discusison}
\subsection{Sustainability Topics within SE Curriculum}
The coverage of the five topics across the five dimensions of the Sustainability Awareness Framework (Table~\ref{tab:sustainability_dimensions}) within the sample program of SE modules is summarized in Table~\ref{tab:modules_vs_sustainability_dimensions}.

\begin{table}[h]
\centering
\caption{Sustainability Dimensions with Related Topics per Modules of a sample Software Engineering Program}
\label{tab:modules_vs_sustainability_dimensions}
\begin{tabular}{p{0.09\textwidth} p{0.35\textwidth}}
\toprule
\textbf{Dimension} & \textbf{Topics} \\
\midrule
Social & Sense of community (HCI, InS, SE, OST, ID, CSec); Trust (SB, HCI, InS, SE, SC, CSec); Inclusiveness \& Diversity (HCI, InS, ADL, FCS, ML, SC, DS); Equality (HCI, InS, ADL, FCS, ML, SC); Participation \& Communication (HCI, SE, SC) \\
\midrule
Individual & \textcolor{gray}{Health}; \textcolor{gray}{Lifelong learning}; Privacy (FCS, InS, HCI, SB, SE); Safety (SE, SB); Agency (HCI, SE) \\
\midrule
Environmental & \textcolor{gray}{Materials and Resources (SC); Soil, Atmospheric and Water Pollution (SC); Biodiversity and Land Use (SC); Logistics and Transportation (SC)}; Energy (SC, IP, PiC, ID, SE)  \\
\midrule
Economic & Value (SD, InS, SC); Customer Relationship Management (SE, HCI, InS); Supply chain (CSec, FCS, SC); Governance and Processes (FCS, OST); Innovation and R\&D (SE, ID, InS, HCI) \\
\midrule
Technical & Maintainability (OOP, SE, CSys); Usability (HCI, InS, ID, SE); Extensibility (OOP, SE) and Adaptability (OOP, SE); Security (CSec,  FCS, SB); Scalability (SE, OST)\\
\bottomrule
\end{tabular}
\end{table}

From this we observe that the topics of the Environmental and Individual dimensions are rather sparsely addressed in our reviewed curriculum. All topics of Environmental dimension, except for Energy, are left out from all modules, except for Sustainable Computing. However, SC is an elective module with between 10 and 30 students choosing to take it every year. Thus, at present the vast majority of the software engineering graduates move out into the SE profession without any appreciation of the link between their daily work and its environmental impact. Neither do they consider such aspects of Individual dimension as health and lifelong learning, and safety aspects are discussed only with respect to  security practices. 

While the topic of Energy is addressed in different modules, these are primarily focused on energy performance of algorithms (IP, SE, PiC) or battery life of devices (ID), without linking the energy use to greenhouse gas emissions or any other environmental impact. 
 
Thus, integrating consideration of the Environmental and Individual dimensions is the first step in embedding a holistic notion of sustainability into the SE curriculum.

\subsection{Sustainability Integration Strategies} \label{sec:sustIntegrationStrategies}

\subsubsection{From Adding-On to Redefining Curriculum}
Based on a review of current approaches to integrating new knowledge and skills (from employability to sustainability) into the engineering curriculum, Kolmos et al  \cite{kolmos2016response} suggest use of three distinct strategies (see Table \ref{tab:educationalFrameworks}), each of which corresponds to a different potential learning outcome (as per Sterling et al \cite{sterling2001sustainable,sterling2011transformative}, see Table \ref{tab:educationalFrameworks}): 

\begin{table}[htbp]
\centering
\caption{Sustainability Integration Strategies and Respective Learning Potential.}
\begin{tabular}{p{0.24\textwidth}|p{0.2\textwidth}}
\hline
\textbf{Strategies \cite{kolmos2016response}} & 
\textbf{Learning Potential \cite{sterling2001sustainable}} \\ \hline

\textit{Add-on strategy:} Small changes, such as incorporating new courses into a largely unchanged curriculum. &
\textit{Adaptive:} Focusing on cognitive aspects. No involvement or questioning of current values or habits. \\ \hline

\textit{Integration strategy:} Changes are promoted at the unit level to embed competency activities. The strategy demands more coordination between units and throughout the curriculum. &
\textit{Critically reflective adaptation:} Teaching specific `sustainability' values or skills that are already known. \\ \hline

\textit{Re-building strategy:} Blending disciplines and social context through effective coordination and management of the curriculum. Priority is given to ensuring values change. &
\textit{Transformative:} Learning and sustainability attributes are constantly reframed through a `creative, reflexive, and participative process'. \\ \hline
\end{tabular}
\label{tab:educationalFrameworks}
\end{table}

\textit{Add-on} is the first strategy, which does not require any change to the existing curriculum, but simply adds a new module to the existing program (e.g, the Sustainable Computing module in our example university) \cite{kolmos2016response}. This could also include other co-curricular activities both credit bearing and not. For instance, in the University of Bristol the the CS school run sustainability-themed hackathons, and offered sustainability champion positions through the Student Union to help engage the students with this topic. This strategy results in adding sustainability ideas to the overall system, without generating significant change \cite{sterling2001sustainable}. Though, this strategy help those students who take such modules or engage with co-/extra-curricular activities to ``do things better" \cite{sterling2011transformative}, they do not change the overall education or respective professional practice.

\textit{Integration} strategy engages with a wider set of units in a given program, integration changes teaching and learning activities so as the sustainability-specific competences can be embedded within all units \cite{kolmos2016response}. This is the main activity undertaken through our curriculum review, aiming to coordinate and integrate the teaching, learning, and assessment for accounting sustainability engineering as a professional competence of software engineering graduates. Sterling at al. \cite{sterling2001sustainable} suggest that this approach will help to re-orient the current system and incorporate sustainability ideas permanently into the  educational system. The students will graduate with sustainability engineering skills and values. Or, in other words, they will learn to `do better things' \cite{sterling2011transformative}. 

\textit{Re-building} strategy is suggested as the most effective one, aiming to blend the disciplines and completely re-define the engineering curriculum \cite{kolmos2016response}. Through such re-built curriculum students would learn to constantly re-frame their understanding of sustainability through creative and reflective practice. In other words, they will learn to `see things differently' \cite{sterling2011transformative}, e.g., considering their roles as software engineers not as developers of products and services, but as constructors of the foundational infrastructure in accordance with which society and businesses operate. Appreciation of the criticality of the role that software and its engineers play in engineering sustainability into the modern world would likely require re-building of not only SE curriculum, but also that of other disciplines, blending many of them together to emerge as new professions  (e.g., including Software, Environmental Science, Business and Social Sciences, including Psychology and Sociology, to name a few). While such review and re-building of the curricula is yet to come, we advocate the need of  immediate assimilation of the integration strategy, as a necessary and urgent step forward.  

\subsubsection{On Practical Path}
Considering the practical steps of the  \textit{integration} strategy,  earlier, we noted that following the interviews, academics agreed on several concrete steps to strengthen the coverage of sustainability concerns within their units. We attribute the relative ease of agreement that the academics showed in this process to a number of factors, which could be of practical relevance to others as well. These factors, we believe are centered around
\textit{creating the common vision on sustainability} through an artifact-supported (i.e., via the SAF dimensions and questions alongside the Green Foundation’s patterns) and student-driven process.

As suggested by Besterfield-Sacre et al. \cite{besterfield2014changing}, having a common vision facilitates the agreement on what the curriculum should be. In our case: 
\begin{enumerate}
    \item The common view on \textit{what sustainability is} was explained through the SAF framework's dimensions, topics and questions;
    \item \textit{How it can be directly related to the taught content} of each module was exemplified with the Green Software Foundation patterns;
    \item The interest expressed by the student body in having sustainability integrated into the teaching content re-enforced the \textit{relevance} of the proposed initiative. 
    \item Finally, the consultative, iterative, and open approach to the module review \textit{respected academic independence} and allowed for step-wise integration, reducing academics' resistance to change.
\end{enumerate}

\section{Conclusion}\label{sec:conclusion}

This paper has examined how sustainability is currently addressed within a sample SE curriculum at the University of Bristol, UK. We showed that two strategies are already in use: an add-on approach (e.g., introducing a Sustainable Computing module and extra/co-curricular activities) and an integration approach (systematically reviewing each module to identify and embed relevant sustainability concerns). We also outlined several practical steps to support others seeking to replicate the integration strategy in their own programs. These include: adopting a clear reference framework such as SAF \cite{duboc2020requirements}; providing concrete, domain-relevant examples such as GSF patterns \cite{greenswpatterns2025}; engaging students and industrial/advisory boards to demonstrate relevance; and running consultative, iterative processes that preserve academic freedom while fostering long-term academic engagement.

We also discussed the potential future role of the rebuilding strategy within our discipline. However, we argue that the integration strategy is a necessary and urgent prerequisite. Only by completing integration can we effectively prepare software engineering graduates as sustainability-aware professionals.

It is important to distinguish the distinct goals of the three strategies. The add-on strategy creates a visible response to sustainability but largely preserves the status quo, leaving SE practice and education unchanged. The integration strategy, by contrast, re-orients SE curricula to embed sustainability knowledge and skills, ensuring graduates are equipped to “do better things” as professionals. We see this as an essential step for the discipline to meaningfully engage with sustainability—the most pressing societal challenge of our time. Finally, the rebuilding strategy seeks to renegotiate the role and boundaries of SE within a changing world. While this may not yet be an immediate pressure, it is approaching, and the SE community would need to begin considering it.

\begin{acks}
Acknowledgments:
The authors would like to thank Henna Parmar - the Sustainability Student Champion for Computer Science at UoB, who carried out the data collection for this project.
\end{acks}

\bibliographystyle{ACM-Reference-Format}
\bibliography{bibliography}

\appendix

\end{document}